\documentstyle[epsf,aps,prl,twocolumn]{revtex}
\def\la{\lambda} 
\def\ep{\epsilon}

\def\om{\omega}

\def\dl{\delta} 
 
\def\si{\sigma} 

\def\nn{\nonumber} 
\def\beqn{\begin{equation}} 
\def\eeqn{\end{equation}}
\def\beqna{\begin{eqnarray}} 
\def\eeqna{\end{eqnarray}}
\def\bea{\begin{array}} 
\def\ea{\end{array}} 
\def\o{\over} 

\def\dg{\dagger} 
\begin{document} 
\title{ A Hopf bifurcation with a strong incommensurate frequency response in
 quantum wells} 
\author{Adriano A. Batista$^1$,  Bjorn Birnir$^2$ and
Pablo I. Tamborenea$^3$}
\address{ $^1$Department of Physics,University of California, Santa
Barbara, CA 93106 } 
\address{ $^2$Mathematics Department, University of
California, Santa Barbara, CA 93106\\ and Science Institute, University of
Iceland, 107 Reykjavik, Iceland}
\address{ $^3$Departamento de Fisica, FCEN,
Universidad de Buenos Aires,
(1428) Buenos Aires, Argentina}
\date{\today}
 \maketitle 
\begin{abstract}
The behavior of delta-doped wide quantum well heterostructures in the
presence of intense far-infrared radiation is studied using semiconductor 
Bloch equations. A quantum well is designed where 
one can either obtain a strong subharmonic (period-doubling) or a strong
incommensurate (Hopf) frequency response by varying the sheet density and 
field strength. These strong responses are easily attainable with current quantum 
well
technology and the field amplitudes and frequencies of the drive are well 
within the range of the free electron laser. 
\end{abstract}
 The effective description of some quantum mechanical systems, far from a 
classical correspondence, can be nonlinear.
Strongly driven THz intersubband transitions in
AlGaAs/GaAs quantum wells are a good example of such a system and
have been shown to exhibit novel nonlinear
phenomena. Many-body effects which cause the nonlinearities become much
more prominent in wide quantum wells ($\approx 300$\AA) where intersubband
spacing is of the order of 10meV which is approximately of the order of
the electron-electron Coulomb interactions. The main signature of the
nonlinearities is the dynamic shielding by the electron gas of the
incoming radiation. The screening blue shifts the absorption peak
frequency, by an amount proportional to the intersubband population difference and sheet density,
to the dressed frequency at which collective oscillations of the entire
electron gas occur. This depolarization shift also causes the generation 
of second-harmonics of 
the drive frequency when it is at half of this dressed
frequency \cite{Hey94}. The
depolarization shift and superharmonic generation have been accurately
simulated and measured, see\cite{Gal96a,Hey94,Hey95}. In addition to this 
dressing of the
intersubband frequency there is a dynamic Stark effect 
which shifts the absorption peak  by an amount proportional to the intensity of the incoming radiation.
Experiments by Heyman et al.\cite{Hey94} and Craig et al.\cite{crai96}
showed the intersubband absorption peak distorting and shifting to the red
towards the bare intersubband frequency as it saturates with increasing
THz intensity. These results were in good agreement with a two-subband
density matrix model proposed by Zalu\.zny\cite{Zal93a,Zal93b}. 

 A superharmonic
generation is a general feature of classical nonlinear systems with
periodic driving, whereas a subharmonic generation or an incommensurate 
frequency response are not as ubiquitous and
usually require the fine tuning of parameters. The same could be the case
in (effectively nonlinear) quantum mechanical systems and this
necessitates a theoretical search for the appropriate parameter ranges. The
generic bifurcation from a stationary solution or a fixed point, corresponding
to a periodic orbit, in a classical system is a
Hopf bifurcation~\cite{RT} and it typically leads to a response comparable in
magnitude to the drive. This
motivates the search for a strong response (a Hopf bifurcation from a fixed
point) in our effectively nonlinear
quantum mechanical system that is a second frequency response of the
collective electron excitations incommensurate with the frequency of the
laser drive. Theoretically this response can be explained by a parametric
oscillation of the electron gas where the system is able to respond at 
a new frequency that is a dressing (as explained above) of the frequency
$E_1-E_0 = E_{10}$ \cite{bat01}.  

Another objective of this letter is to verify the
predictions of the model of Galdrikian and Birnir\cite{Gal96a} for period
doubling bifurcation for a more realistic situation in the quantum well.
The present model is more complete and makes less approximations on the
electron-electron Coulomb interaction and also includes the exchange terms
in the Hamiltonian. On the other hand analytical calculations as the
averaging technique used in \cite{bat00} are
considerably more difficult to perform now. Therefore we are limited to
focus on numerical results for the nonlinear dynamics.

	Sherwin et al.\cite{she95,she98} have studied the optical properties of wide
quantum wells of $Al_{0.3}Ga_{0.7}As$ (barrier) and GaAs (well) driven by
the far infrared radiation of the free electron laser (FEL)  at UCSB. The epitaxially grown structures are filled with a sheet
density of $\approx 1.5\times 10^{11}e/ cm^{2}$ provided by silicon donor layers located
outside of the well. The donors are usually at hundreds of \AA's away to avoid
scattering of the electrons by the dopants, thus increasing the
coherent radiation output of the electrons collective oscillations.

	Many-body effects on intersubband transitions and optical
properties of QWs have been studied mainly with Hartree-Fock and rotating wave
approximations for the response of two-(sub)band QWs \cite{haug90,chu92,luo93}. Based on
similar methods Nikonov, Imamoglu et al.\cite{nik97,nik99} demonstrated
that the influence of exchange and depolarization terms on the absorption
or gain spectrum disappears in the limit of very narrow quantum wells.
 Numerical integration of the multisubband semiconductor
Bloch equations were performed by Tsang et al\cite{tsa92}, although they
neglected depolarization shift terms, which is a serious handicap for their
predictions for wide quantum wells. 

In this letter we study the semiconductor Bloch
equations without RWA from a nonlinear dynamical systems perspective and
find, in qualitative agreement with previous work by Birnir and
Galdrikian\cite{Gal96a}, that for the two-subband driven near resonance
period doubling bifurcations or optical bistability may occur. For a
three-subband QW designed for this project and driven at resonance 
$\om=E_{20}$ several types of
bifurcations may occur depending on the shape of the well and the sheet density.
We observe a Hopf bifurcation producing a strong signal at a lower frequency
than the frequency of the drive. This signal corresponds to a dressing of
$E_{10}$ and for a different (lower) value of the field strength a
period-doubling bifurcation is observed that also produces a strong
signal but now at half the frequency of the drive. Both signals are produced
at moderate values of the field strength. The latter result is a big improvement
over the result in \cite{Gal96a} that was much weaker than the fundamental and occured
at relatively large field strength. Our approach can in
principle be applied to any multisubband systems, but in practice when $\om\geq E_{20}$ one 
may run in problems such as coupling the system to the continuum 
or the excitation of LO
phonons (if $\om>36$meV). However, these problems can be avoided by
introducing more barriers in the well which create several closely spaced 
subbands as in Figure 1. 
   
The optical properties of the confined electrons in
the QW can be studied with the many-body Hamiltonian written in terms of field operators.
The electron contribution is given by 
{\small 
\beqna \hat H_{el}&=& \int
d^3x
\hat\psi^\dg(x,t)(-{1\o2m}\hbar^2\nabla^2+W(z))\hat\psi(x,t)+\nn\\&&
{e^2\o2}\int d^3x\int
d^3x'\hat\psi^\dg(x,t)\hat\psi^\dg(x',t){1\o|x-x'|}\hat\psi(x',t)\hat\psi(x,t)\nn\\&&-e{\cal E}(t)\int d^3x\hat\psi^\dg(x,t)z\hat\psi(x,t), 
\eeqna}where $W(z)$ is the bare QW potential. It is also assumed that the electric
field generated from the radiation off the collective oscillations inside
the QW are negligible compared with the laser light ${\cal E}$. The field 
operators 
expressed as a linear combination of annihilation
 operators $a_{k,n}$ are 
 \[ \hat\psi(x,t)\equiv\sum_{k,n}{e^{ik\cdot\rho}\o
\sqrt{A}}\xi_n(z)a_{k,n}(t) 
\]
where $\xi_n(z)$ is the self-consistent
n-th Hartree eigenstate, $k$ is the in-plane wave vector and $A$ is the QW area.
The Hamiltonian can be rewritten in terms of (time-dependent) annihilation and creation operators as 
{\small 
\beqna 
H&=&\sum _k E_n(k)a_{k,n}^\dagger a_{k,n}-{1\o 2}{\sum}
_{k,k',\{n\}}
V^0_{\{n\}}a^\dg_{k,n_1}a^\dg_{k',n_2}a_{k',n_3}a_{k,n_4}\nn\\&&+{1\o
2}{\sum} _{q\neq0,k,k',\{n\}}
V_{\{n\}}(q)a^\dg_{k+q,n_1}a^\dg_{k'-q,n_2}a_{k',n_3}a_{k,n_4}\nn\\&&+\sum_{k,k',n_1,n_2,n}V^0_{n_1nnn_2}f_n(k')a_{k,n_1}^\dagger
a_{k,n_2} \nn\\&&-\sum_{k,n,m} \mu_{n,m}(k){\mathcal E}(t)a_{k,n}^\dagger a_{k,m},
\label{ham3} 
\eeqna
}where $E_n(k)$ is the self-consistent Hartree
energy, $\mu_{n,m}$ is the dipole moment between the n-th and m-th states, $f_n(k)$ is the Fermi occupation number for the nth-subband and
the Coulomb interaction terms are 
{\small 
\[  V_{\{n\}}(q)={2\pi e^2\o
\ep_0A}\int dz_1 \int dz_2
{\tiny \frac{\xi_{n_1}(z_1)\xi_{n_2}(z_2)\xi_{n_3}(z_2) \xi_{n_4}(z_1)}{q}
e^{-q|z_1-z_2|}}
\]
\[ V^0_{\{n\}}={2\pi e^2\o \ep_0A}\int dz_1\int dz_2
\xi_{n_1}(z_1)\xi_{n_2}(z_2)|z_1-z_2|\xi_{n_3}(z_2)\xi_{n_4}(z_1) 
\]
}where $\ep_0$ is the static dielectric constant in GaAs, which is approximately 13. In
addition to the usual exchange terms that appear in the electron gas
Hamiltonian we obtain depolarization shift terms in the Hamiltonian which
are due to the electron shielding and the quasi two-dimensional character of the problem (it is not
negligible in wide quantum wells with sheet densities $ >10^{11}e/\mbox{cm}^{2}$).
The interaction with the laser light is introduced through the electric
dipole approximation which implies that any momentum transfer due to the
THz radiation is neglected. 

Hamiltonian in hand, we proceed with the
derivation of the equations of motion for the density matrix in the time-dependent
Hartree-Fock approximation, which are also known as the semiconductor Bloch
equations. They provide the collective optical response of the electrons
to the laser light. In order to be more realistic we introduce dissipation
in our equations of motion. Energy (inelastic) and momentum (elastic)
scatterings are accounted for albeit phenomenologically (by the
relaxation-time approximation). 

 The dynamics for the density matrix elements
$\si_{n,n'}(k)=~<a^\dg_{k,n}a_{k,n'}> $ are obtained from the Heisenberg
equations of motion for the operator $a^\dg_{p,n}a_{p,n'}$ followed by a
quantum statistical average and the Hartree-Fock approximation.  We
finally obtain, see \cite{bat01} for details, 
{\small
\beqna
&-&i\hbar\dot\si_{n,n'}(k)=E_{n,n'}(k)\si_{n,n'}(k)+i{\hbar\o T_{nn'}}\big(\si_{n,n'}(k)-f_n(k)\dl_{nn'}\big)\nn\\
&-&\sum_{q\neq 0,n_1n_2n_3}V_{n_1n_2n_3n}(q)\si_{n_1n_3}(|k-q|)\si_{n_2n'}(k)\nn\\&+&\sum_{q\neq
0,n_2n_3n_4}V_{n'n_2n_3n_4}(q)\si_{nn_3}(k)\si_{n_2n_4}(|k-q|)\nn\\&-&\sum_{k',n_2n_3n_4}V^0_{n'n_2n_3n_4}\si_{n,n_4}(k)\big(\si_{n_2n_3}(k')-f_{n_2}(k')\dl_{n_2,n_3}\big)\nn\\
&-&\sum_{k',n_1n_2n_3}V^0_{n_1n_2n_3n}\si_{n_1,n'}(k)\big(\si_{n_2n_3}(k')-f_{n_2}(k')\dl_{n_2,n_3}\big)\nn\\&+&\sum_m{\mathcal E}(t)\big(\mu_{mn}\si_{mn'}(k)-\mu_{n'm}\si_{nm}(k)\big)
\label{heis1} 
\eeqna 
}where $E_{n,n'}(k)= E_n(k)-E_{n'}(k)$ and $T_{nn'}$ is the experimentally
measured relaxation time, see Heymann et al.\cite{Hey95}. Another approximation is to consider the
effective masses of all subbands equal, which is fairly accurate
for AlGaAs QW's. We also neglect the exchange
terms since their inclusion does not modify qualitatively our results and the analysis becomes simpler. It can be shown analytically that this is a good approximation for high sheet densities, since then the kinetic energy terms dominate \cite{fett71}. With these approximations and a sum over $k$ performed, the dynamics
for the averaged density matrix becomes 
{\small
 \beqna
&-&i\hbar{\partial\o\partial
t}\si_{n,n'}=(E_n-E_{n'})\si_{n,n'}+i{\hbar\o T_{nn'}}(\si_{n,n'}-\si^0_{nn'})\nn\\&+&A
N_s\sum_{n_2n_3n_4}V^0_{n'n_2n_3n_4}\si_{n,n_4}(\si_{n_2n_3}-\si^0_{n_2n_3})\nn\\
&&-AN_s\sum_{n_1n_2n_3}V^0_{n_1n_2n_3n}\si_{n_1,n'}(\si_{n_2n_3}-\si^0_{n_2n_3})
\nn\\&&-\sum_m{\mathcal E}(t)\big(\mu_{mn}\si_{mn'}-\mu_{n'm}\si_{nm}\big) 
\label{heis2} 
\eeqna 
}where $\si_{n,n'}={1\o AN_s}\sum_k \si_{n,n'}(k)$, $N_s$ is the sheet
density and $\si^0_{nn'}$ is the equilibrium density matrix. The equations
were put in dimensionless form by dividing all terms by $E_{20}$.

\begin{figure}[tb]
\epsfxsize=2.5in \centerline{\epsffile{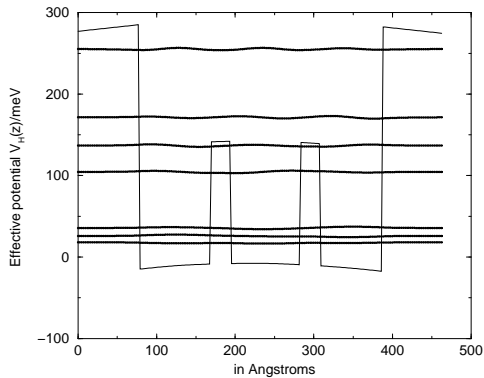}} \caption{The
stationary self-consistent potential for a sheet density of $N_s=3.0\times 10^{11}e/ \mbox{cm}^{2}$ , and its first seven eigenstates
with asymptotes set to their energies. The double barrier creates, through tunneling, three closely spaced subbands well isolated
from upper subbands. The slight asymmetry enhances the
nonlinear effects, but too much asymmetry reduces the tunneling and 
with it the decoupling with upper subbands.}
\label{fig:well}
\end{figure}

 The QW studied was designed to have length 310 \AA, depth of 300 meV, two barrier
steps of 26\AA~in width and 150meV in height. Figure \ref{fig:well} 
shows the effective well-shape 
with a sheet
density $N_s=3.0\times10^{11}e/\mbox{cm}^{2}$. The equations (\ref{heis2}) 
for three subbands only were
integrated using the fourth-order Runge-Kutta method with 2048 steps per
cycle of drive and the results are plotted in Figures 2-4.

\begin{figure}[tb] 
\epsfxsize=2.5in \centerline{\epsffile{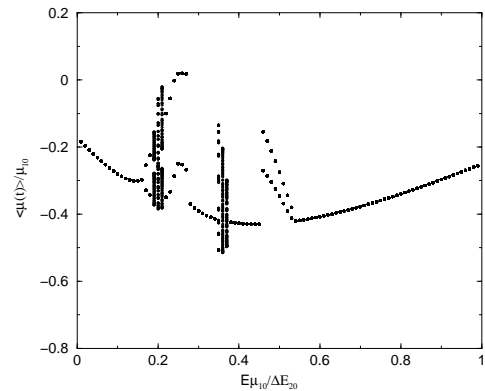}}
\caption{ The Poincar\'e map of the averaged dipole moment in the doped 
QW with a sheet density of $N_s=3.0\times 10^{11}e/\mbox{cm}^{2}$. With 
increasing field strength a PDB occurs at $\la=0.17$. At $\la=0.378$ 
a supercritical Hopf bifurcation occurs, as the field is decreased below this
value. The two branches of the PDB undergo a Hopf bifurcation at $\la=0.19$.
} 
\label{fig:dm} 
\end{figure}

\begin{figure}[tb]
\epsfxsize=2.5in \centerline{\epsffile{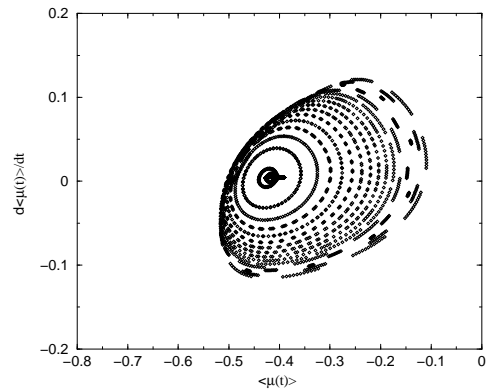}} \caption{
Phase portrait view of the Hopf bifurcation in the quantum well. Each one of the closed orbits corresponds to a different value of the field amplitude, with $\la=0.34-0.38$ and a sheet density of $N_s=3.0\times 10^{11}e/\mbox{cm}^{2}$. 
} 
\label{fig:hopf}
\end{figure}

 Figure \ref{fig:dm} shows the bifurcation diagram of the dipole moment of
the collective oscillations. The scaled dipole moment $<\mu(t)>/\mu_{10}$ is
plotted as a function of the unitless field amplitude 
$\la=e{\mathcal E}z_{10}/E_{20}$. A PDB in the Poincar\'e map 
at $\la=0.17$ followed by a Hopf bifurcation (as the 
field decreases) at 
$\la=0.378$ are observed. We found an optimum sheet density in which the 
bifurcations occur with greatest strength and at the same time require
small field amplitude, such information can be used as a guide for the 
experimentalist. For $N_s<0.5\times10^{11}e/\mbox{cm}^{2}$ and for $N_s>6.0\times 10^{11}e/\mbox{cm}^{2}$ the bifurcations 
become too small for practical observations. 
The phase portrait of the
Poincar\'e map in Figure \ref{fig:hopf} shows the scaled dipole moment
plotted against its time derivative (on the vertical axis) for various
values of field strength before and after the Hopf bifurcation. This plot is
the strongest evidence that Hopf bifurcations occur for our system. 
Near $\la=0.378$ a supercritical
Hopf bifurcation is observed as we decrease the field intensity, 
the fixed point, a periodic orbit with the frequency of
the drive, becomes unstable generating a small closed orbit. Several of
those orbits are plotted each one corresponding to a fixed value of the
field strength, decreasing the field amplitude gives a new and larger
closed orbit. 

The period-doubling can be understood as a two-photon process, see \cite{Gal96a}, its increased strength in this letter is due to the
fact that it is being assisted by the presence of almost equally spaced (in frequency) frequencies $E_{10}$ and $E_{20}$. The dynamics of the electron gas 
correspond to a periodic orbit of twice the period of the response before
the bifurcation.

\begin{figure}[tb] \epsfxsize=2.5in
\centerline{\epsffile{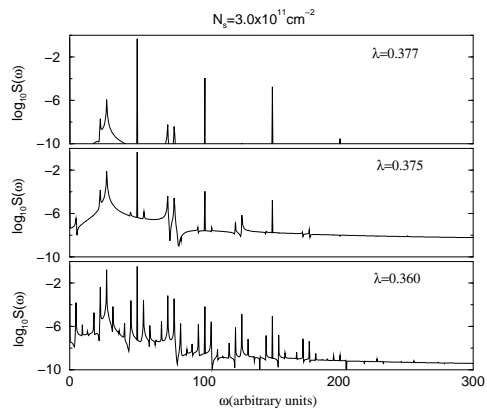}} \caption{Power spectra
of the dipole moment for the doped asymmetric quantum well near the
onset of the supercritical Hopf bifurcation. As we decrease the field past $\la=0.38$ the low frequncy signal grows continuously. The zero frequency peak coincides with
the y-axis and extends to zero.} \label{fig:ps} \end{figure}

The large and broad peak at roughly $\om/2$ is the main
feature of the
bifurcation as can be seen in Figure \ref{fig:ps} which shows the
power spectrum before and after the Hopf bifurcation. The peak in Figure
\ref{fig:ps} is broad due to the
difference frequencies, dressed $\pm (E_{21}-E_{10})$, that are clearly visible
on each side of the incommensurate peak. The half-frequency peaks (not shown) for
the PDB are sharp in distinction\cite{Gal96a}. By varying the sheet density and field strength
the incommensurate peak can be tuned but this will
be explored in Batista, Tamborenea and Birnir\cite{bat01}. In Figure 4 we 
also observe a DC response that concides with the frequency axis. The 
dynamics of the electron gas after (decreasing $\lambda$) the Hopf bifurcation
corresponds to a quasi periodic orbit on a torus defined by the incommensurate
frequency and the fundamental. At both the PDB and Hopf bifurcations the 
populations of the upper subbands are suddenly increased at the
expense of the zeroth subband. The frequency of the drive is
of the order of 20 meV, which is in the range of the FEL  
(however, $E_{20}$ can be lowered by decreasing the tunneling if necessary). 

The realization that QWs 
can undergo
a Hopf bifurcation (from a periodic orbit) only observed before in classical 
nonlinear systems can
lead to many applications including frequency down-converters. In our QW the
lower subbands are almost equally spaced, but that could be easily changed
by increasing its asymmetry and eventually eliminating the possibility of
PDBs. The asymmetry though cannot be too big  otherwise the tunnelling is
reduced and we cannot guarantee that our three-subband system driven
at resonance with $E_{20}$ is really isolated from upper subbands and the
continuum. 

The authors would like to thank Mark Sherwin, S. James Allen and Atac Imamoglu 
for helpful comments and
discussions. The idea of nonlinear bifurcations in quantum 
wells were inspired by J. D.
Crawford's elegant exposition on bifurcation
theory\cite{craw91}. The work was supported by NSF grants DMS-9704874,
DMS-0072191 and a DARPA-Navy grant N00014-99-10935.

\begin{thebibliography}{10}
\bibitem{RT}
The Hopf bifurcation is well known to be related to the Ruelle-Takens
  quasi-periodic route to chaos in classical systems \cite{craw91}.
\end{thebibliography}
\end{document}